\documentclass[prb,preprint]{revtex4-2} 
\usepackage{amsfonts} % needed for bold Greek, Fraktur, and blackboard bold
\usepackage{graphicx} % needed for figures
\usepackage{amsmath} %math symbols package
\usepackage{amssymb} %math symbols package
\usepackage{amsthm} %math symbols package
\usepackage{bm} %bold math symbols package
\usepackage{color} %colored text package
\usepackage{amstext}
\usepackage{hyperref}

\bibpunct{}{}{,}{s}{}{\textsuperscript{,}}

\begin{document}

\title{Making squeezed-coherent states concrete by determining their wavefunction}

\author{E.~Mungu\'ia-Gonz\'alez}
%email{elihumunguia7@gmail.com} % optional
\affiliation{Facultad de Ciencias, Universidad de Colima, Bernal Diaz del Castillo 340, 28045, Colima, Mexico}
\author{S.~Rego}
%\email{regosheldon2003@gmail.com}
\affiliation{R. N. Podar School (CBSE), Jain Derasar Marg, Santacruz West, Mumbai - 400054, India}
%\altaffiliation[permanent address: ]{Department of Physics, Georgetown University, 37th and O Sts. NW, Washington, DC 20057 USA} % optional second address
% If there were a second author at the same address, we would put another 
\author{J.~K.~Freericks} 
%\email{james.freericks@georgetown.edu}
% \author statement.
\affiliation{Department of Physics, Georgetown University, 37th and O Sts. NW, Washington, DC 20057 USA}
% Please provide a full mailing address here.

\date{\today}

\begin{abstract}
With the successes of { the Laser Interferometer Gravitational-wave Observatory}, we anticipate increased interest in working with squeezed states in the undergraduate and graduate quantum-mechanics classroom. Because squeezed-coherent states are minimum uncertainty states, their wavefunctions in position and momentum space must be Gaussians. But this result is rarely discussed in treatments of squeezed states in quantum textbooks or quantum optics textbooks. In this work, we show three different ways to construct the wavefunction for squeezed-coherent states: (i) a differential equation-based approach; (ii) an approach that uses an expansion in terms of the simple-harmonic oscillator wavefunctions; and (iii) a fully operator-based approach. We do this to illustrate that the concept of the wavefunction can be introduced no matter what methodology an instructor wishes to use. We hope that working with the wavefunction will help demystify the concept of a squeezed-coherent state.
\end{abstract}
% AJP requires an abstract for all regular article submissions.
% Abstracts are optional for submissions to the "Notes and Discussions" section.

\maketitle % title page is now complete

\section{Introduction} % Section titles are automatically converted to all-caps.
% Section numbering is automatic.

{ The Laser Interferometer Gravitational-wave Observatory} (LIGO) amazed the scientific world with its first observation of a black-hole merger in 2015\cite{LIGO_exp}. { LIGO is a modern engineering marvel that is designed to detect gravitational waves through their differential action on massive mirrors that are placed at the ends of two 4 km-long arms of a Michelson-Morley interferometer (supplemented by Fabry-Perot multireflection cavities); signal discrimination is achieved by requiring the gravitational-wave signal to be observed at two separate installations separated by thousands of kms.} This measurement apparatus measures distances to an amazing one part in $10^{21}$. { The sensitivity of LIGO is mainly limited by vacuum fluctuations in the electromagnetic field. These enter the interferometer through the unused input port of the beam splitter (that is, the side of the beam-splitter that is not illuminated by the laser). LIGO's sensitivity is then improved by inputting a squeezed-vacuum state} on the normally empty input of the beam splitter of the Michelson interferometer~\cite{caves1,caves2}. { Electromagnetic fields can be measured by an amplitude quadrature (measuring the amplitude of the field) or by measuring a phase quadrature (measuring where the field goes through zero). These are conjugate measurements, which satisfy uncertainty relations that have the same magnitude of uncertainty for each quadrature in the vacuum state. The squeezed-vacuum state is the ground state of an effective harmonic oscillator with a Gaussian waveform that has a different width than the ground state has. This initiates a time-varying quantum state whose width varies in time from smaller to larger and back again. By timing measurements to when the uncertainty for the measured quadrature is smaller than in a coherent state, one can perform measurements with higher precision than the standard quantum limit.}

This remarkable achievement motivates the idea of adding (or discussing in more detail) squeezed states within the quantum mechanics curriculum. While we do not have any specific data on student opinions, it seems reasonable to assume that the topic of squeezed states is more difficult for students than that of coherent states. Coherent states can be described simply in terms of a translation of position or momentum of the ground state. But a squeezed state changes the shape of the { ground-state} wavefunction (it remains Gaussian, but with complex parameters that are changed), which we believe is a more difficult concept to master. 

The focus of this work is to describe the wavefunctions themselves and not to be a complete review on squeezed states. We will also briefly describe how a squeezed-vacuum state is used in upgraded LIGO measurements. There already exist a number of pedagogical reviews of other properties of squeezed states, so we do not repeat that work here. In particular, we refer the reader to the wonderful introduction to squeezed states by Henry and Glotzer\cite{henry_glotzer}, which provides a comprehensive introduction to squeezed states (but does not discuss the wavefunctions) and to the classic by Loudon and Knight\cite{loudon_knight}, which emphasizes phase-space pictures for the time evolution of uncertainty in squeezed-coherent states and how one actually creates squeezed states of light for use in quantum optics experiments.

% While we do not have any specific data on student opinions, it seems to be reasonable to assume that the topic of squeezed states is more difficult for students than coherent states---coherent states can be described simply in terms of a translation (of position and momentum) of the ground state, while a squeezed state changes the shape of the wavefunction (it remains Gaussian, but with complex parameters that are changed). We believe this is a more difficult concept to master.

%A brief history of squeezed states and squeezed coherent states

We start with a brief review of the history of coherent and squeezed states; for more details, see the review of Dodunov.\cite{dodunov} Coherent states were introduced by Schrödinger\cite{schroedinger_1926} in 1926 and squeezed states by Kennard\cite{kennard} in 1927. The wavefunction for squeezed states was written down by Plebanski\cite{plebanski} in 1956. But  the full potential of these states was brought to the fore by the work of Glauber\cite{glauber} in the 1960s. When squeezed states of light were ultimately observed in 1985 by Slusher\cite{slusher} and collaborators, the years of theoretical developments were finally realized in the lab. Today, squeezed states are widely used within quantum sensing and routinely improve the sensitivity of a range of different sensors, including LIGO. Hence, it is increasingly important for students to learn about coherent states, {squeezed-vacuum states} and squeezed-coherent states. The squeezed-coherent state is the most general and we focus on it in this work. However, one can simplify many of the derivations below to just { the case of} squeezed-vacuum states, if desired.

%A description of why a minimum uncertainty state must have a Gaussian wavefunction

Because students often visualize quantum concepts by reasoning through wavefunctions,\cite{hiller,visual_qm,per_wavefunctions} it is useful to represent these states in terms of position-space or momentum-space wavefunctions.  Indeed, because squeezed-coherent states are minimal uncertainty states for real values of the squeezing parameter ({  introduced below}), { their wavefunctions} must be { real-valued} Gaussians in those cases. The proof of this is well-known\cite{mus}.  We briefly repeat it next.

The standard Robertson approach to the uncertainty principle~\cite{robertson} relates the uncertainty of two   self-adjoint operators $\hat{A}$ and $\hat{B}$ to the expectation value of their commutator $\hat{C}=[\hat{A},\hat{B}]$. Using the Schwarz inequality, which bounds the modulus square of the inner product of two vectors by the product of their norms squared, we have
\begin{equation}
    |\langle \psi|(\hat{A}-a)(\hat{B}-b)|\psi\rangle|^2\le \langle \psi|(\hat{A}-a)^2|\psi\rangle\langle\psi|(\hat{B}-b)^2|\psi\rangle.\label{eq:schwartz}
\end{equation}
Here, $a=\langle\psi|\hat{A}|\psi\rangle$ and $b=\langle\psi|\hat{B}|\psi\rangle$ and both  {  expectation values over the states $|\psi\rangle$} are real. Using the fact that the commutator of two Hermitian operators is anti-Hermitian, we choose to minimize the modulus squared on the left-hand side via the imaginary part of the inner product
\begin{equation}
    \big (\text{Im}\langle \psi|(\hat{A}-a)(\hat{B}-b)|\psi\rangle \big )^2\le\big |\langle \psi|(\hat{A}-a)^2|\psi\rangle\big|\,\big|\langle\psi|(\hat{B}-b)^2|\psi\rangle\big |.\label{eq:unc1}
\end{equation}
We note that Hermiticity of the operators allows us to simplify the imaginary part via
\begin{eqnarray}
   \big (\text{Im}\langle \psi|(\hat{A}-a)(\hat{B}-b)|\psi\rangle \big )^2&=&\left (\frac{\langle \psi|(\hat{A}-a)(\hat{B}-b)|\psi\rangle-\langle \psi|(\hat{A}-a)(\hat{B}-b)|\psi\rangle^*}{2i}\right )^2\\
   &=&\left (\frac{\langle \psi|(\hat{A}-a)(\hat{B}-b)|\psi\rangle-\langle \psi|(\hat{B}-b)(\hat{A}-a)|\psi\rangle}{2i}\right )^2\\
   &=&\left (\frac{\langle\psi|[\hat{A},\hat{B}]|\psi\rangle}{2i}\right )^2=-\frac{1}{4}\big (\langle\psi|\hat{C}|\psi\rangle\big )^2.\label{eq:unc2}
\end{eqnarray}
For  position and momentum { operators, we have} $[\hat{x},\hat{p}]=i\hbar$, { and hence}
\begin{equation}
    \sqrt{\langle\psi|\hat{x}^2|\psi\rangle-\langle\psi|\hat{x}|\psi\rangle^2}\sqrt{\langle\psi|\hat{p}^2|\psi\rangle-\langle\psi|\hat{p}|\psi\rangle^2}\ge \frac{\hbar}{2}.\label{eq:unc_xp}
\end{equation}
The minimum uncertainty state saturates the inequality bound. Thinking in terms of the inner product between vectors, equality with the product of the norms in Eq.~(\ref{eq:unc1}) means that both vectors are parallel (or that  $(\hat{x}-\langle\psi|\hat{x}|\psi\rangle)|\psi\rangle\propto (\hat{p}-\langle\psi |\hat{p}|\psi\rangle)|\psi\rangle$), and the inner product must be purely imaginary to saturate the bound. We define $x_0=\langle\psi|\hat{x}|\psi\rangle$ and $p_0=\langle\psi|\hat{p}|\psi\rangle$ and take the product of the previous equality with the position bra $\langle x|$ to find that
\begin{equation}
    \left (-i\hbar\frac{d}{dx}-p_0\right )\psi(x)=i\lambda(x-x_0)\psi(x),\label{eq:mus1}
\end{equation}
with $\lambda$ a real { dimensionful} constant. Solving, we find that
\begin{equation}
    \psi(x)=\gamma e^{-\frac{\lambda}{\hbar}\left (\frac{1}{2}x^2-x_0x\right )+\frac{i}{\hbar}p_0x}\label{eq:mus2}
\end{equation}
where $\gamma$ is a { dimensionful} complex number coming from the constant of integration. This proves that all minimal uncertainty states must be Gaussians. But, the converse is not true! If the Gaussian { form for the wavefunction} has a complex coefficient $\lambda$, it will not be minimal uncertainty. For the minimal uncertainty Gaussians, the free parameters are the normalization constant, a multiplicative complex phase, the { real} width of the Gaussian, and the complex number $\frac{\lambda}{\hbar} x_0+\frac{i}{\hbar}p_0$ multiplying the $x$ in the exponent. The main point of this work is to determine these parameters for a general squeezed-coherent state { (where $\lambda$ can be complex)}; we will find they are not always minimal uncertainty states, but they still assume the above Gaussian form { of the wavefunction} in Eq.~(\ref{eq:mus2}).

%Describe why it is hard to see this simple behavior

Why is this not routinely discussed in textbooks? Well, we often work with squeezed-coherent states in terms of their operators and in this case, the Gaussian form of the wavefunction is hidden (discussed in Sec.~IIB below). As we will see, it is not so easy to extract it from this operator form, hence it often is not covered in textbooks (one advanced textbook that does cover it via the differential equation approach is Barnett and Radmore\cite{barnett_radmore}). We believe { this topic} should be covered in the classroom, because this simple picture relating the Gaussian to the minimal uncertainty state is so compelling; it allows students to connect squeezed-coherent states with the more familiar Gaussian functions. Is it { more relevant} to discuss squeezed-coherent states, or just  squeezed-vacuum states? We choose to discuss the most general case because it is a straightforward exercise to simplify this work { for} {squeezed-vacuum} states, if desired. Squeezed-coherent states with a fixed squeezing parameter provide an overcomplete basis, just as coherent states do, and they have been used in three recent experiments. The first is in improving stimulated emission microscopy for biological systems,\cite{microscope} enabling the imaging of biological structures that do not fluoresce. { The} sensitivity { of such microscopes} is limited by the quantum fluctuations in the probe beam; using a squeezed-coherent probe beam improves performance. Squeezed-coherent states { can also be used} as a way to boost the detection of small displacements by moving them outside of the zero-point-motion fluctuations, { so they can be more easily detected}. This is achieved by first squeezing the vacuum, then having the displacement occur, then squeezing the state in an orthogonal direction. 
%This process of squeezing, displacing and squeezing in an orthogonal direction, amplifies the displacement, so it lies outside the vacuum fluctuations and can be more easily detected. 
This has been used in detecting the displacement of ions in an ion trap,\cite{wineland} and as a way to measure the small changes induced by axions in resonant-microwave-cavity-based searches for dark matter.\cite{caves_axion1,caves_axion2}
    
%A brief derivation of a squeezed coherent state.

We end this introduction with a ``crash course'' in squeezed-coherent states to set our notation. Coherent states are created by simply applying a simultaneous translation in position and momentum to the ground state of the simple harmonic oscillator. The simple harmonic oscillator Hamiltonian for an oscillator with a mass $m$ and frequency $\omega$ is
\begin{equation}
    \hat{H}=\frac{\hat{p}^2}{2m}+\frac{1}{2}m\omega^2\hat{x}^2=\hbar\omega\left (\hat{a}^\dagger\hat{a}+\frac{1}{2}\right ).\label{eq:Ham}
\end{equation}
The raising and lowering operators are $\hat{a}=\sqrt{\frac{m\omega}{2\hbar}}\hat{x}+i\frac{1}{\sqrt{2m\omega\hbar}}\hat{p}$ and $\hat{a}^\dagger=(\hat{a})^\dagger$, which satisfy $[\hat{a},\hat{a}^\dagger]=1$.
The simultaneous translation operator is called the displacement operator and is given by 
\begin{equation}
    e^{\frac{i}{\hbar}(p_0\hat{x}-x_0\hat{p})}=e^{\alpha\hat{a}^\dagger-\alpha^*\hat{a}}=\hat{D}(\alpha),\label{eq:displacement}
\end{equation}
with $\alpha=\sqrt{\frac{m\omega}{2\hbar}}x_0+i\frac{1}{\sqrt{2m\omega\hbar}}p_0$. { It corresponds to translating the ground state by $x_0$ and $p_0$, respectively.} The coherent state is defined by
\begin{equation}
    |\alpha\rangle=\hat{D}(\alpha)|0\rangle,\label{eq:coherent_state}
\end{equation}
and is an eigenstate of the lowering operator, satisfying $\hat{a}|\alpha\rangle=\alpha|\alpha\rangle$.

Squeezed states are similarly defined, but with a quadratic in the exponent. In this work we only need a simplified squeezed state, which is all we define and work with. We start by introducing { the complex} parameter $\xi=re^{i\phi}$, { where} $r$ and $\phi$ are real. The squeezing operator $\hat{S}(\xi)$ is defined by
\begin{equation}
    \hat{S}(\xi)=e^{-\frac{\xi}{2} (\hat{a}^\dagger)^2+\frac{\xi^*}{2}\hat{a}^2}\label{eq:squeezing}
\end{equation}
and is a unitary operator  because $\hat{S}^\dagger(\xi)=\hat{S}(-\xi)=\hat{S}^{-1}(\xi)$. Then, the squeezed-coherent state { is formed by applying} the squeezing operator { onto the ground state}, followed by { applying} the displacement operator, or $\hat{D}(\alpha)\hat{S}(\xi)|0\rangle$. Generalizations { in which} operators are applied to excited states can be found in the literature, but { here} we always focus on operators being applied to the simple-harmonic-oscillator ground state $|0\rangle$. Note that while it is true that the effect of the displacement operator on the wavefunction can be incorporated afterwards, 
%by working just with the squeezed states and then translating position and momentum (and adding in a phase factor) afterwards, 
we choose to work with the squeezed-coherent states directly because we feel that students often view those methods as ``trick solutions'' that they cannot master themselves.

\section{Three ways to compute the squeezed-state wavefunction}

%The wavefunction in the literature, especially Shumaker and the Danes. 
The wavefunction of the squeezed-coherent state appears in the literature in a number of places. The two most prominent ones are the 1986 review by Bonny Shumaker~\cite{shumaker} and the 1996 article by M\o ller, J\o rgensen and Dahl~\cite{moller}. We follow closely the Shumaker approach for the differential-equation-based strategy; we adapt the M\o ller approach for the simple-harmonic-oscillator-based method, and we provide a third method that uses only operators, in the spirit of Mu\~noz-Tapia\cite{munoz_tapia} and Rushka and Freericks.\cite{rushka} { After completing this work, we were made aware of related work on the technique of integration within an ordered product of operators, which could be extended to provide a fourth method to calculate squeezed-coherent state wavefunctions,\cite{iwop} but we do not discuss that method further here.}

\subsection{Differential-equation-based approach}

%Use the annihilation operator definition to compute this. Discuss how to find the overall constant.
The differential-equation-based approach starts from the defining relation of the ground state,  $\hat{a}|0\rangle=0$, which says the lowering operator annihilates the ground state. To determine how this is modified for the squeezed-coherent state, we multiply on the left by the squeezing operator followed by the displacement operator, and multiply by 1 between the ground state and the annihilation operator, to yield
\begin{equation}
    0=\hat{a}|0\rangle=\hat{D}(\alpha)\hat{S}(\xi)\hat{a}\hat{S}(-\xi)\hat{D}(-\alpha)\underbrace{\hat{D}(\alpha)\hat{S}(\xi)|0\rangle}_{\text{squeezed-coherent state}},\label{eq:de1}
\end{equation}
where we used the facts that { $\hat{D}^{-1}(\alpha)=\hat{D}(-\alpha)$ and $\hat{S}^{-1}(\xi)=\hat{S}(-\xi)$}. The underbrace highlights the appearance of the squeezed-coherent state. One can see that the lowering operator has two nested similarity transformations and can be evaluated by applying the Hadamard lemma twice;  see Eq.~(\ref{eq:hadamard}). When we apply the Hadamard { lemma} with the squeezing operator, we find a repeating infinite series given by
\begin{equation}
    \hat{S}(\xi)\hat{a}\hat{S}(-\xi)=\sum_{n=0}^\infty \frac{1}{(2n)!}|\xi|^{2n}\hat{a}+\sum_{n=0}^\infty\frac{1}{(2n+1)!}|\xi|^{2n}\xi\hat{a}^\dagger,
\end{equation}
which is found by simply calculating all of the nested commutators in the Hadamard lemma and collecting the even and odd terms. The first summation is $\cosh |\xi|$, while the second is $\frac{\sinh|\xi|}{|\xi|}$. Then the Hadamard { lemma applied to $\hat{D}(\alpha)\hat{a}\hat{D}(-\alpha)$ (and its Hermitian conjugate)}  truncates after the first commutator (shifting the $\hat{a}$ by $-\alpha$ and the $\hat{a}^\dagger$ by $-\alpha^*$).
We end up with
\begin{equation}
    \left (\frac{\xi}{|\xi|}\sinh(|\xi|)(\hat{a}^\dagger-\alpha^*)+\cosh(|\xi|)(\hat{a}-\alpha)\right )\hat{D}(\alpha)\hat{S}(\xi)|0\rangle=0.\label{eq:de2}
\end{equation}
We construct the differential equation in a few steps: (i) we multiply from the left by $\langle x|$ and define the wavefunction by $\psi_{\xi,\alpha}(x)=\langle x|\hat{D}(\alpha)\hat{S}(\xi)|0\rangle$; (ii) we replace the raising and lowering operators by the position and momentum operators according to their definitions; (iii) we evaluate the position operator on the position eigenstate yielding the position eigenvalue and we replace the momentum operator by $-i\hbar\frac{d}{dx}$ in the standard way; and (iv) we replace $|\xi|=r$ and $\xi/|\xi|=e^{i\phi}$.

In order to avoid confusion with typesetting in equations, we note that every hyperbolic sine, cosine and tangent that appears in equations will have $r$ as its argument.
This then gives the linear first-order differential equation
\begin{equation}
   \frac{d}{dx}\psi_{\xi,\alpha}(x)=-\left \{\frac{m\omega}{\hbar}\,\frac{\cosh r+e^{i\phi}\sinh r}{\cosh r-e^{i\phi}\sinh r}(x-x_0)-\frac{ip_0}{\hbar}\right \}\psi_{\xi,\alpha}(x),\label{eq:de3}
\end{equation}
which can be immediately integrated to give
\begin{equation}
    \psi_{\xi,\alpha}(x)=c\exp\left (-\frac{\cosh r+e^{i\phi}\sinh r}{\cosh r -e^{i\phi}\sinh r}\,\frac{m\omega}{2\hbar}(x-x_0)^2+\frac{ip_0 x}{\hbar}\right ). \label{eq:de_integrated}
\end{equation}
The challenge is in determining the constant $c$. The magnitude of $c$ is a normalization factor and then there is a remaining multiplicative complex phase factor. But, even now, one sees that the Gaussian form is verified from the differential equation calculation.

The calculation of the constant phase $c$ uses two steps. First, we take the overlap of the squeezed-coherent state with the harmonic oscillator ground state.
This requires working with some additional operator identities (the Weyl form of the Baker-Campbell-Hausdorff identity in Eq.~(\ref{eq:BCH}) and the exponential disentangling identity in Eq.~(\ref{eq:disentangling2}) in the appendix). Both are standard identities. We proceed as follows:
\begin{eqnarray}
\langle 0|\hat{D}(\alpha)\hat{S}(\xi)|0\rangle&=&\langle 0|e^{\alpha\hat{a}^\dagger-\alpha^*\hat{a}}e^{-\frac{1}{2}\xi(\hat{a}^\dagger)^2+\frac{1}{2}\xi^*\hat{a}^2}|0\rangle\label{eq:constant2}\\
&=&\langle 0|e^{\alpha\hat{a}^\dagger}e^{-\alpha^*\hat{a}}e^{-\frac12|\alpha|^2}e^{-\frac{1}{2}e^{i\phi}\tanh r(\hat{a}^\dagger)^2}e^{-\frac{1}{2}\ln(\cosh r) (\hat{a}^\dagger\hat{a}+\hat{a}\hat{a}^\dagger)}e^{\frac{1}{2}e^{-i\phi}\tanh r\hat{a}^2}|0\rangle\label{eq:constant3}\\
&=&e^{-\frac12|\alpha|^2}\frac{1}{\sqrt{\cosh r}}\langle 0|e^{-\alpha^*\hat{a}}e^{-\frac{1}{2}e^{i\phi}\tanh r(\hat{a}^\dagger)^2}e^{\alpha^*\hat{a}}|0\rangle\label{eq:constant4}\\
&=&e^{-\frac12|\alpha|^2}\frac{1}{\sqrt{\cosh r}}\langle 0|e^{-\frac{1}{2}e^{i\phi}\tanh r(\hat{a}^\dagger-\alpha^*)^2}|0\rangle\label{eq:constant5}\\
&=&e^{-\frac12\left (|\alpha|^2+e^{i\phi}\tanh r\alpha^{*2}\right )}\frac{1}{\sqrt{\cosh r}}\label{eq:constant5a}\\
&=&\frac{\exp\left (-\frac{m\omega}{4\hbar}(1+e^{i\phi}\tanh r)x_0^2+i\frac{x_0p_0}{2\hbar}e^{i\phi}\tanh r-\frac{p_0^2}{4m\omega\hbar}(1-e^{i\phi}\tanh r)\right )}{\sqrt{\cosh r}},\label{eq:constant6}
\end{eqnarray}
where in the {first} line, we used the definition of the operators and in the { second}, we used the Weyl form of the Baker-Campbell-Hausdorff identity in Eq.~(\ref{eq:BCH}) and the exponential disentangling identity from Eq.~(\ref{eq:disentangling2}). In the {third} line, we used the facts that the lowering operator annihilates the ground state ket, and the raising operator annihilates the ground-state bra. In the { fourth}, we { employ} the Hadamard { lemma} using Eq.~(\ref{eq:hadamard}). The {fifth} line follows from the fact that the ground state is normalized, $\langle 0|0\rangle =1$. {The last line re-expresses the result in terms of the other set of variables.}

We can { also} use the fact that the position eigenstates form a complete set to  evaluate this expectation value via
\begin{eqnarray}
    &~&\langle 0|\hat{D}(\alpha)\hat{S}(\xi)|0\rangle =\int_{-\infty}^\infty dx\,\langle 0|x\rangle\,\langle x|\hat{D}(\alpha)\hat{S}(\xi)|0\rangle\label{eq:constant7}\\
    &~&~~~~=c\left (\frac{m\omega}{\pi\hbar}\right )^{\frac14}\int_{-\infty}^\infty dx\,e^{-\frac{m\omega}{2\hbar}x^2}\exp\left (-\frac{\cosh r+e^{i\phi}\sinh r}{\cosh r -e^{i\phi}\sinh r}\,\frac{m\omega}{2\hbar}(x-x_0)^2+\frac{ip_0 x}{\hbar}\right )\label{eq:constant8}\\
    &~&~~~~=c\left (\frac{\pi\hbar}{m\omega}\right )^{\frac14}\sqrt{\frac{\cosh r -e^{i\phi}\sinh r}{\cosh r}}\exp\left (-\frac{m\omega}{4\hbar}(1+e^{i\phi}\tanh r)x_0^2+i\frac{x_0p_0}{2\hbar}(1+e^{i\phi}\tanh r)\right )\nonumber\\
    &~&~~~~~~~~~\times \exp\left (-\frac{p_0^2}{4m\omega\hbar}(1-e^{i\phi}\tanh r)\right ),\label{eq:constant9}
\end{eqnarray}
where we used the fact that the normalized ground state wavefunction is $\langle x|0\rangle=\left (\frac{m\omega}{\pi\hbar}\right )^{\frac14}e^{-\frac{m\omega}{2\hbar}x^2}$ and the explicit result in Eq.~(\ref{eq:de_integrated}). The integral in the second line is a Gaussian, which is easier to integrate after shifting $x\to x+x_0$, and then yields Eq.~(\ref{eq:constant9}). Equating Eq.~(\ref{eq:constant9}) with Eq.~(\ref{eq:constant6}) gives us

\begin{eqnarray}
    c&=&e^{-\frac{i}{2\hbar}x_0p_0}\frac{1}{\sqrt{\cosh r -e^{i\phi}\sinh r}}\left (\frac{m\omega}{\pi\hbar}\right )^{\frac{1}{4}}\\
    &=&e^{-\frac{i}{2\hbar}x_0p_0}\frac{1}{\sqrt{|\cosh r -e^{i\phi}\sinh r|}}\frac{\sqrt{\cosh r -e^{-i\phi}\sinh r}}{\sqrt{|\cosh r -e^{-i\phi}\sinh r|}}\left (\frac{m\omega}{\pi\hbar}\right )^{\frac{1}{4}},\label{eq:constant_final}
\end{eqnarray}
where in the second line we rationalized the denominator to separate out the normalization factor (second term) and the multiplicative complex phase (third term), in the form Shumaker\cite{shumaker} uses.
Putting this all together, with Eq.~(\ref{eq:de_integrated}), gives us the wavefunction in position space
\begin{eqnarray}
\psi_{\xi,\alpha}(x)&=&\left (\frac{m\omega}{\pi\hbar}\right )^{\frac{1}{4}}\frac{1}{\sqrt{|\cosh r-e^{i\phi}\sinh r|}}\frac{\sqrt{\cosh r-e^{-i\phi}\sinh r}}{\sqrt{|\cosh r-e^{-i\phi}\sinh r|}}e^{-\frac{i}{2\hbar}x_0p_0}e^{\frac{i}{\hbar}p_0x}\nonumber\\&\times&\exp\left (-\frac{\cosh r+e^{i\phi}\sinh r}{\cosh r -e^{i\phi}\sinh r}\,\frac{m\omega}{2\hbar}(x-x_0)^2\right ),\label{eq:wf_x_final}
\end{eqnarray}
which is, of course, a Gaussian; it is a minimal uncertainty state when $\phi=0$ or $\phi=\pi$ (since that yields a real coefficient to the $x^2$ term).
Following similar steps to compute the wavefunction in momentum space gives us the following result:
\begin{eqnarray}
\tilde{\psi}_{\xi, \alpha}(p)&=& \left(\frac{1}{\pi m\omega\hbar} \right)^{\frac{1}{4}} \frac{1}{\sqrt{|\cosh r + e^{i\phi} \sinh r|}}\frac{\sqrt{\cosh r + e^{-i\phi} \sinh r}}{\sqrt{|\cosh r + e^{-i\phi} \sinh r|}}e^{\frac{i}{2\hbar}x_0p_0}e^{-\frac{i}{\hbar}x_0 p}\nonumber\\&\times&\exp\left(- \frac{\cosh r - e^{i\phi}\sinh r}{\cosh r + e^{i\phi}\sinh r}\,\frac{1}{2m\omega\hbar} (p - p_0)^2 \right).\label{eq:wf_p_final}
\end{eqnarray} 
These results agree exactly with those given by Shumaker\cite{shumaker}.

Since we have explicit expressions for the wavefunctions in position and momentum space, we can use them to immediately determine the fluctuations of position and momentum, which yield
\begin{equation}
    \sqrt{\langle \psi_{\xi,\alpha}|\hat{x}^2|\psi_{\xi,\alpha}\rangle-\langle \psi_{\xi,\alpha}|\hat{x}|\psi_{\xi,\alpha}\rangle^2}=\sqrt{\frac{\hbar}{2m\omega}}\sqrt{\cosh(2r) - \sinh(2r) \cos(\phi)}\label{eq:x_unc}
\end{equation}
and
\begin{equation}
    \sqrt{\langle \psi_{\xi,\alpha}|\hat{p}^2|\psi_{\xi,\alpha}\rangle-\langle \psi_{\xi,\alpha}|\hat{p}|\psi_{\xi,\alpha}\rangle^2}=\sqrt{\frac{\hbar m\omega}{2}}\sqrt{\cosh(2r) + \sinh(2r) \cos(\phi)}.\label{eq:p_unc}
\end{equation}
We do not show the details of these calculations because they are long and this {  approach} is not the most efficient way to compute them. Note that when $\phi=0$, the position uncertainty is reduced by a factor of $e^{-r}$ while the momentum uncertainty is enhanced by $e^r$ so that the uncertainty product is minimal; the opposite occurs for $\phi=\pi$.

\subsection{Simple-harmonic-oscillator-expansion-based approach}

We now discuss how one can determine the Gaussian wavefunction from the harmonic oscillator wavefunctions. We are influenced by the authoritative work by M\o ller, J\o rgensen and Dahl\cite{moller} in this derivation. We will find that this approach  is most appropriate for treatments where the squeezed states are introduced via operator manipulations, as found in some undergraduate textbooks\cite{puri,steeb} and in MIT's undergraduate MOOC by Barton Zwiebach\cite{mooc}. It requires the use of a highly nontrivial identity that relates sums of Laguerre polynomials to confluent hypergeometric functions  and so may not be appropriate for undergraduate classes.

In this approach, we simply multiply the squeezed state by a position bra $\langle x|$ so that $\psi_{\xi,\alpha}(x)=\langle x|\hat{D}(\alpha)\hat{S}(\xi)|0\rangle$. Using a similar operator methodology as the one we used to lead up to Eq.~(\ref{eq:constant3}), but using position and momentum operators for the displacement operator instead of the raising and lowering operators, gives us
\begin{equation}
    \psi_{\xi,\alpha}(x)=\langle x|e^{\frac{i}{\hbar}p_0\hat{x}}e^{-\frac{i}{\hbar}x_0\hat{p}}e^{-\frac{i}{2\hbar}x_0p_0}e^{-\frac{1}{2}e^{i\phi}\tanh r(\hat{a}^\dagger)^2}e^{-\frac{1}{2}\ln(\cosh r) (\hat{a}^\dagger\hat{a}+\hat{a}\hat{a}^\dagger)}e^{\frac{1}{2}e^{-i\phi}\tanh r\hat{a}^2}|0\rangle.\label{eq:showf1}
\end{equation}
The $\hat{x}$ operator can be immediately evaluated on the bra, since it is an eigenstate, then the $\hat{p}$ operator translates the position eigenstate by $-x_0$. Similarly, the { the ground state is annihilated by $\hat{a}$ and is an eigenstate with respect to $\hat{a}^\dagger\hat{a}+\hat{a}\hat{a}^\dagger$}. This then  gives us
\begin{equation}
    \psi_{\xi,\alpha}(x)=e^{\frac{i}{\hbar}p_0x-\frac{i}{2\hbar}p_0x_0}\frac{1}{\sqrt{\cosh r}}\langle x{-}x_0|e^{-\frac{1}{2}e^{i\phi}\tanh r(\hat{a}^\dagger)^2}|0\rangle.\label{eq:showf2}
\end{equation}
Expanding the exponential in a power series we recognize that $\frac{1}{\sqrt{(2n)!}}\langle x-x_0|(\hat{a}^\dagger)^{2n}|0\rangle$  is the coordinate-space wavefunction of the 2n-th excited state, which is expressed in terms of Hermite polynomials, so that 
\begin{eqnarray}
\psi_{\xi,\alpha}(x)&=&\left (\frac{m\omega}{\pi\hbar}\right )^{\frac14} e^{-\frac{m\omega}{2\hbar}(x-x_0)^2+\frac{i}{\hbar}p_0x-\frac{i}{2\hbar}p_0x_0}\frac{1}{\sqrt{\cosh r}}\nonumber\\
&\times&\sum_{n=0}^\infty \frac{(-1)^n}{n!}\left (\frac{1}{4}e^{i\phi}\tanh r\right )^{n}H_{2n}\left (\sqrt{\frac{m\omega}{\hbar}}(x-x_0)\right ).\label{eq:showf3}
\end{eqnarray}
In the general case, one might stop here, { however, seeing that $\psi_{\xi,\alpha}(x)$ has a Gaussian form} appears to be completely opaque { at this stage}. To see how this is a Gaussian requires one to use a little-known identity, found by Erd\'elyi, relating Laguerre polynomials to the confluent hypergeometric function\cite{bateman}
\begin{equation}
    _1F_1\left (a;c;\frac{xy}{x-1}\right )=(1-x)^a\sum_{n=0}^\infty \frac{(a)^{(n)}}{(c)^{(n)}}L_n^{(c-1)}(y)x^n,\label{bateman}
\end{equation}
where $_1F_1(a;b;x)$ is the confluent hypergeometric function of the first kind (also called $M$).
Next, using the fact that Hermite polynomials of an even index can be written in terms of Laguerre polynomials as
\begin{equation}
    H_{2n}(x)=(-1)^n2^{2n}n!L_n^{(-\frac{1}{2})}(x^2),\label{eq:laguerre-hermite}
\end{equation}
we can rewrite the summation in Eq.~(\ref{eq:showf3}) as
\begin{eqnarray}
    &~&\sum_{n=0}^\infty \frac{(-1)^n}{n!}\left (\frac{1}{4}e^{i\phi}\tanh r\right )^{n}H_{2n}\left (\sqrt{\frac{m\omega}{\hbar}}(x-x_0)\right )\nonumber\\
    &~&~~~~~~~~~~~~~=\sum_{n=0}^\infty L_n^{(-\frac{1}{2})}\left (\frac{m\omega}{\hbar}(x-x_0)^2\right ) \left (e^{i\phi}\tanh r\right )^{n}\nonumber\\
    &~&~~~~~~~~~~~~~=\frac{1}{\sqrt{1-e^{i\phi}\tanh r}}\,_1F_1\left (\frac{1}{2};\frac{1}{2};-\frac{m\omega(x-x_0)^2e^{i\phi}\tanh r}{\hbar(1-e^{i\phi}\tanh r)}\right )\nonumber\\
    &~&~~~~~~~~~~~~~=\frac{1}{\sqrt{1-e^{i\phi}\tanh r}}\exp\left (-\frac{2e^{i\phi}\sinh r}{\cosh r-e^{i\phi}\sinh r}\,\frac{m\omega}{2\hbar}(x-x_0)^2\right ),\label{eq:summation}
\end{eqnarray}
because $_1F_1(a;a;x)=e^{x}$.
Substituting this into Eq.~(\ref{eq:showf3}) then gives us the final result, which is exactly the result in Eq.~(\ref{eq:wf_x_final}). Obviously, one can also go through this approach in momentum space and this yields the result in Eq.~(\ref{eq:wf_p_final}), as it must.

While this approach focuses directly on expanding in terms of the familiar simple-harmonic-oscillator eigenfunctions, it requires the use of an identity that is unlikely to have been seen by any student (and probably most faculty, as well). This makes the approach challenging to carry out, without spending time discussing the required identity in some detail. 

%Use the expansion based on completeness of the SHO wavefunction to do this. Our approach will be operator-based, but it can be done other ways as well. 

%Discuss the identity needed to obtain the Gaussian.

\subsection{Full-operator-based approach}

This section is inspired by the operator-based approach described by Mu\~noz-Tapia\cite{munoz_tapia}, but generalized to include exponential operator identities (as developed in Ref.~\onlinecite{rushka}) that were not used in Mu\~noz-Tapia.
We require three operator identities (the Weyl form of the Baker-Campbell-Hausdorff formula, the Hadamard lemma, and the exponential disentangling identity). As we show below, one must be proficient in  working with these identities to complete the calculation. We summarize all of the required mathematical preliminaries in the Appendix.

%Derive the position and momentum states in terms of squeezed states. Do we discuss denseness of the SHO states here?

To start this calculation, we first want to express the position and momentum states in terms of an operator acting on the simple-harmonic-oscillator ground state. This is possible because, when the squeezing parameter approaches $\pm\infty$, the transformed lowering operator becomes proportional to the position or momentum operator. Hence the squeezed-coherent state eigenfunction relation becomes that of the position or momentum eigenstate. We explain this in detail next.

The starting point is Eq.~(\ref{eq:de2}) with $\phi=0$. We multiply it by $\sqrt{\frac{\hbar}{2m\omega\cosh r}}$
and use the exponential disentangling identity { from Eq.~(\ref{eq:disentangling2})} to yield
\begin{eqnarray}
&~&\sqrt{\frac{\hbar}{2m\omega\cosh r}}\hat{D}(\alpha)\underbrace{\hat{D}(-\alpha)(\sinh r(\hat{a}^\dagger-\alpha^*)+\cosh r(\hat{a}-\alpha))\hat{D}(\alpha)}_{\text{ simplifiable via the Hadamard lemma}}\nonumber\\
&~&~~~~\times e^{-\frac{1}{2}\tanh r(\hat{a}^\dagger)^2}e^{-\frac{1}{2}\ln(\cosh r) (\hat{a}^\dagger\hat{a}+\hat{a}\hat{a}^\dagger)}e^{\frac{1}{2}\tanh r\hat{a}^2}|0\rangle=0\label{eq:squeeze1}\\
&~&\sqrt{\frac{\hbar}{2m\omega\cosh r}}\hat{D}(\alpha)(\sinh r\,\hat{a}^\dagger+\cosh r\,\hat{a}) e^{-\frac{1}{2}\tanh r(\hat{a}^\dagger)^2}\frac{1}{\sqrt{\cosh r}}|0\rangle=0\label{eq:squeeze2}\\
&~&\sqrt{\frac{\hbar}{2m\omega}}(\hat{a}+\tanh r\hat{a}^\dagger)\underbrace{e^{-\frac{1}{2}\tanh r(\hat{a}^\dagger)^2}|0\rangle}_{\text{state}}=0.\label{eq:squeeze_eig_eq}
\end{eqnarray}
Here, we used the facts that annihilation operators destroy $|0\rangle$ and the Hadamard { lemma} in Eq.~(\ref{eq:hadamard}) removes the shifts by $\alpha$ and $\alpha^*$. We drop the displacement operator on the left hand side in the third line, because it is unitary, and hence cannot make the terms to the right of it vanish; hence they must vanish without it.

If we take the limit of $r\to\infty$, the leftmost operator becomes \(\hat{x}\). This means we have a state (indicated by the underbrace after the limit is taken) that is annihilated by the position operator. This is then proportional to the position eigenstate at the origin. Hence $|x{=}0\rangle\propto e^{-\frac{1}{2}(\hat{a}^\dagger)^2}|0\rangle$. Using the translation operator, we find that 
\begin{equation}
    |x\rangle=c^\prime e^{-\frac{i}{\hbar}x\hat{p}}e^{-\frac{1}{2}(\hat{a}^\dagger)^2}|0\rangle.\label{eq:pos_eigstate}
\end{equation}
To determine the normalization $c^\prime$, we multiply by the simple-harmonic-oscillator ground-state bra from the left and find
\begin{equation}
    \psi_{gs}^*(x)=\langle 0|x\rangle=c^\prime\langle 0|e^{-\frac{i}{\hbar}x\hat{p}}e^{-\frac{1}{2}(\hat{a}^\dagger)^2}|0\rangle.\label{eq:norm1}
\end{equation}
We write the momentum operator in terms of the ladder operators and use the Weyl form of the Baker-Campbell-Hausdorff formula in Eq.~(\ref{eq:BCH}) with the raising operator to the left  to yield
\begin{equation}
    \langle 0|x\rangle=c^\prime\langle 0|e^{\sqrt{\frac{m\omega}{2\hbar}}x\hat{a}^\dagger}e^{-\sqrt{\frac{m\omega}{2\hbar}}x\hat{a}}e^{-\frac{m\omega}{4\hbar}x^2}e^{-\frac12(\hat{a}^\dagger)^2}|0\rangle
\end{equation}
Next, we operate the exponential of the raising operator on the ground-state bra and multiply by one on the ground-state ket to { use a Hadamard lemma}; we also introduce the ground-state wavefunction on the left-hand side. This produces 
\begin{equation}
    \left (\frac{m\omega}{\pi\hbar}\right )^{\frac{1}{4}}e^{-\frac{m\omega}{2\hbar}x^2}=c^\prime e^{-\frac{m\omega}{4\hbar}x^2}\langle 0|e^{-\sqrt{\frac{m\omega}{2\hbar}}x\hat{a}}e^{-\frac{1}{2}(\hat{a}^\dagger)^2}e^{\sqrt{\frac{m\omega}{2\hbar}}x\hat{a}}|0\rangle.\label{eq:norm2}
\end{equation}
Now, we evaluate the Hadamard lemma using Eq.~(\ref{eq:hadamard})
\begin{equation}
    \left (\frac{m\omega}{\pi\hbar}\right )^{\frac{1}{4}}e^{-\frac{m\omega}{4\hbar}x^2}=c^\prime\langle 0|e^{-\frac{1}{2}\left (\hat{a}^\dagger-\sqrt{\frac{m\omega}{2\hbar}}x\right )^2}|0\rangle,\label{eq:norm3}
\end{equation}
and then we annihilate the raising operator on the bra to the left. This tells us $c^\prime=\left (\frac{m\omega}{\pi\hbar}\right )^{\frac{1}{4}}$. Hence,
\begin{equation}
    |x\rangle=\left (\frac{m\omega}{\pi\hbar}\right )^{\frac{1}{4}}e^{-x\sqrt{\frac{m\omega}{2\hbar}}(\hat{a}-\hat{a}^\dagger)}e^{-\frac{1}{2}(\hat{a}^\dagger)^2}|0\rangle.\label{eq:pos_eigenstate_final}
\end{equation}
This result agrees with previous calculations that develop the position eigenstate in terms of operators acting on the ground-state ket\cite{fan,moya-cessa}. 

Of course, one can repeat this for momentum eigenstates as well. The final result is
\begin{equation}
    |p\rangle=\left(\frac{1}{\pi m\omega\hbar}\right)^{\frac{1}{4}}
    e^{\frac{ip}{\sqrt{2m\omega\hbar}}(\hat{a} + \hat{a}^{\dagger})}e^{\frac{1}{2}(\hat{a}^{\dagger})^2}|0\rangle
    \label{eq:mom_eigenstate_final}
\end{equation}
This construction of position and momentum eigenstates is a nice application of the fact that the solutions of the simple harmonic oscillator form a dense set in the Hilbert space; especially if we formulate the squeezed states in terms of linear combinations of the simple harmonic oscillator wavefunctions, as we did in the previous subsection. But as we take a limit point of these linear combinations (as we do to find the position or momentum eigenstates), the limiting function need not be in the Hilbert space. In fact, we know these two eigenfunctions lie in the tempered distributions of the Gelfand triple.\cite{capri}

%Derive the wavefunction, determining all constants.

We are now ready to compute the squeezed-coherent state wavefunction using only operator identities. We start by taking the overlap of the position eigenstate in Eq.~(\ref{eq:pos_eigenstate_final}) with the squeezed-coherent state $\hat{D}(\alpha)\hat{S}(\xi)|0\rangle$:
\begin{equation}
\psi_{\xi,\alpha}(x)=\left (\frac{m\omega}{\pi\hbar}\right )^{\frac{1}{4}}\langle 0|e^{-\frac{1}{2}\hat{a}^2}e^{x\sqrt{\frac{m\omega}{2\hbar}}(\hat{a}-\hat{a}^\dagger)}e^{\alpha\hat{a}^\dagger-\alpha^*\hat{a}}e^{-\frac{1}{2}\xi(\hat{a}^\dagger)^2+\frac{1}{2}\xi^*\hat{a}^2}|0\rangle.\label{eq:psi_op1}
\end{equation}
We use three operator identities in simplifying these equations: (i) the exponential re-ordering identity (based on the Hadamard lemma), where $e^{\hat{A}}f(\hat{B})=f(e^{\hat{A}}Be^{-\hat{A}})e^{\hat{A}}$; (ii) the Weyl form of the Baker-Campbell-Hausdorff formula; and (iii) the exponential disentangling identity (for more details, see the appendix). Since we have many of these manipulations to do, we describe them carefully and carry them out efficiently in the following formulas.

Our first step is to use exponential reordering in Eq.~(\ref{eq:braiding}) on the first two operators yielding 
\begin{equation}
\psi_{\xi,\alpha}(x)=\left (\frac{m\omega}{\pi\hbar}\right )^{\frac{1}{4}}\langle 0|e^{x\sqrt{\frac{m\omega}{2\hbar}}(\hat{a}-\hat{a}^\dagger)}e^{-\frac{1}{2}\left (\hat{a}-x\sqrt{\frac{m\omega}{2\hbar}}\right )^2}e^{\alpha\hat{a}^\dagger-\alpha^*\hat{a}}e^{-\frac{1}{2}\xi(\hat{a}^\dagger)^2+\frac{1}{2}\xi^*\hat{a}^2}|0\rangle.\label{eq:psi_op2}
\end{equation}
Next, we use the Weyl form of Baker-Campbell-Hausdorff  in Eq.~(\ref{eq:BCH}) on the leftmost term with the raising operator to the left, annihilate the raising operator against the ground-state bra, expand the square in the second operator, move the constants out, and combine the terms linear in the lowering operator to the leftmost term:
\begin{equation}
\psi_{\xi,\alpha}(x)=\left (\frac{m\omega}{\pi\hbar}\right )^{\frac{1}{4}}e^{-\frac{m\omega}{2\hbar}x^2}\langle 0|e^{x\sqrt{\frac{2m\omega}{\hbar}}\hat{a}}e^{-\frac{1}{2}\hat{a}^2}e^{\alpha\hat{a}^\dagger-\alpha^*\hat{a}}e^{-\frac{1}{2}\xi(\hat{a}^\dagger)^2+\frac{1}{2}\xi^*\hat{a}^2}|0\rangle.\label{eq:psi_op3}
\end{equation}
Next, we move the displacement operator to the left, through the two operators to the left of it, until it is next to the ground-state bra. This is done by the exponential re-ordering:
\begin{equation}
\psi_{\xi,\alpha}(x)=\left (\frac{m\omega}{\pi\hbar}\right )^{\frac{1}{4}}e^{-\frac{m\omega}{2\hbar}x^2}\langle 0|e^{\alpha\hat{a}^\dagger-\alpha^*\hat{a}}e^{x\sqrt{\frac{2m\omega}{\hbar}}(\hat{a}+\alpha)}e^{-\frac{1}{2}(\hat{a}+\alpha)^2}e^{-\frac{1}{2}\xi(\hat{a}^\dagger)^2+\frac{1}{2}\xi^*\hat{a}^2}|0\rangle.\label{eq:psi_op4}
\end{equation}
We repeat all of the steps leading up to Eq.~(\ref{eq:psi_op3}) on the leftmost displacement operator to obtain
\begin{equation}
\psi_{\xi,\alpha}(x)=\left (\frac{m\omega}{\pi\hbar}\right )^{\frac{1}{4}}e^{-\frac{m\omega}{2\hbar}x^2+\alpha x\sqrt{\frac{2m\omega}{\hbar}}-\frac{1}{2}(|\alpha|^2+\alpha^2)}\langle 0|e^{\left (x\sqrt{\frac{2m\omega}{\hbar}}-2\text{Re}\,\alpha\right )\hat{a}}e^{-\frac{1}{2}\hat{a}^2}e^{-\frac{1}{2}\xi(\hat{a}^\dagger)^2+\frac{1}{2}\xi^*\hat{a}^2}|0\rangle.\label{eq:psi_op5}
\end{equation}
Now, we expand the rightmost operator with the same exponential disentangling identity we have been using (from Eq.~(\ref{eq:disentangling2})). The lowering operator on the far right annihilates against the ground-state ket. The middle term evaluates to $\frac{1}{\sqrt{\cosh r}}$. This yields
\begin{equation}
\psi_{\xi,\alpha}(x)=\left (\frac{m\omega}{\pi\hbar}\right )^{\frac{1}{4}}\frac{e^{-\frac{m\omega}{2\hbar}x^2+\alpha x\sqrt{\frac{2m\omega}{\hbar}}-\frac{1}{2}(|\alpha|^2+\alpha^2)}}{\sqrt{\cosh r}}\langle 0|e^{\left (x\sqrt{\frac{2m\omega}{\hbar}}-2\text{Re}\,\alpha\right )\hat{a}}e^{-\frac{1}{2}\hat{a}^2}e^{-\frac{1}{2}e^{i\phi}\tanh r(\hat{a}^\dagger)^2}|0\rangle.\label{eq:psi_op7}
\end{equation}
Exponential reordering is now applied to move the middle operator to the right, where the lowering operator annihilates against the ground-state ket:
\begin{equation}
\psi_{\xi,\alpha}(x)=\left (\frac{m\omega}{\pi\hbar}\right )^{\frac{1}{4}}\frac{e^{-\frac{m\omega}{2\hbar}x^2+\alpha x\sqrt{\frac{2m\omega}{\hbar}}-\frac{1}{2}(|\alpha|^2+\alpha^2)}}{\sqrt{\cosh r}}\langle 0|e^{\left (x\sqrt{\frac{2m\omega}{\hbar}}-2\text{Re}\,\alpha\right )\hat{a}}e^{-\frac{1}{2}e^{i\phi}\tanh r(\hat{a}^\dagger-\hat{a})^2}|0\rangle.\label{eq:psi_op8}
\end{equation}
The next step is to expand the exponent of the rightmost operator. We now use an exponential disentangling identity on this term. But it is of a different form than what we used before. This identity is worked out in the Appendix in Eq.~(\ref{eq:disentangling6}). Again, the lowering operator term on the right can be replaced by 1 and the middle term can be evaluated because the ground-state ket is an eigenstate of the number operator. Hence, we have
\begin{eqnarray}
\psi_{\xi,\alpha}(x)&=&\left (\frac{m\omega}{\pi\hbar}\right )^{\frac{1}{4}}\frac{e^{-\frac{m\omega}{2\hbar}x^2+\alpha x\sqrt{\frac{2m\omega}{\hbar}}-\frac{1}{2}(|\alpha|^2+\alpha^2)}}{\sqrt{\cosh r}\sqrt{1-e^{i\phi}\tanh r}}\langle 0|e^{\left (x\sqrt{\frac{2m\omega}{\hbar}}-2\text{Re}\,\alpha\right )\hat{a}}\nonumber\\
&\times&\exp\left (-\frac{1}{2}\frac{e^{i\phi}\tanh r}{1-e^{i\phi}\tanh r}(\hat{a}^\dagger)^2\right )|0\rangle.\label{eq:psi_op9}
\end{eqnarray}
Our last step is to use exponential reordering to change the order of the two operators. Then the lowering operator annihilates to the right and the raising operator annihilates to the left. We are left with numbers only. Recalling that the ground state is normalized ($\langle 0|0\rangle=1$), we finally obtain
\begin{eqnarray}
\psi_{\xi,\alpha}(x)&=&\left (\frac{m\omega}{\pi\hbar}\right )^{\frac{1}{4}}\frac{e^{-\frac{m\omega}{2\hbar}x^2+\alpha x\sqrt{\frac{2m\omega}{\hbar}}-\frac{1}{2}(|\alpha|^2+\alpha^2)}}{\sqrt{\cosh r-e^{i\phi}\sinh r}}\nonumber\\
&\times&\exp\left (-\frac{1}{2}\frac{e^{i\phi}\sinh r}{\cosh r-e^{i\phi}\sinh r}\left (x\sqrt{\frac{2m\omega}{\hbar}}-2\text{Re}\,\alpha\right )^2\right ).\label{eq:psi_op10}
\end{eqnarray}
Finally, if we substitute in the value for $\alpha$ and combine terms, we get the result given in Eq.~(\ref{eq:wf_x_final}). Of course, this can also be repeated for the momentum-space wavefunction (it will require use of the identity in Eq.~(\ref{eq:disentangling7}) towards the end of the derivation) and it produces the expected result in Eq.~(\ref{eq:wf_p_final}).

\section{Time evolution of these states}

%Explain how Hadamard gets all of these. Also show figures/animations here if desired.
%Gaussians in free Hamiltonians always expand. In SHO, they remain fixed (coherent states) or oscillate (squeezed states)

In this section, we develop the time evolution { of the squeezed-coherent state} using exponential reordering of the time evolution operator and the operators in the squeezed-coherent state. Namely, we have
\begin{equation}
    \psi_{\xi,\alpha}(x,t)=\langle x|\hat{U}(t,0)\hat{D}(\alpha)\hat{S}(\xi)|0\rangle=\langle x|e^{-\frac{i}{\hbar}\hat{H}t}e^{\alpha\hat{a}^\dagger-\alpha^*\hat{a}}e^{-\frac{1}{2}\xi(\hat{a}^\dagger)^2+\frac{1}{2}\xi^*\hat{a}^2}|0\rangle.
\end{equation}
One can see that exponential reordering in Eq.~(\ref{eq:braiding}) moves the time-evolution operator $e^{-\frac{i}{\hbar}\hat{H}t}$, all the way to the right, where it evaluates to $e^{-\frac{i\omega t}{2}}$ against the ground-state ket. As it moves through the other two operators, it simply takes $\xi\to\xi e^{-2i\omega t}$ and $\alpha\to\alpha e^{-i\omega t}$. This means the time-dependent wavefunction is found by the substitutions $\phi\to\phi-2\omega t$, $x_0\to x_0\cos\omega t + \frac{p_0}{m\omega} \sin\omega t$ and $p_0\to p_0\cos\omega t - m\omega x_0 \sin\omega t$; it also is multiplied by an overall phase factor $e^{-\frac{i\omega t}{2}}$. Given the simple nature of this result, we do not rewrite the full time-dependent wavefunction here. Note that an alternative way to determine time evolution is described by Gersch.\cite{gersch}

\begin{figure}
\includegraphics[width=3.75in]{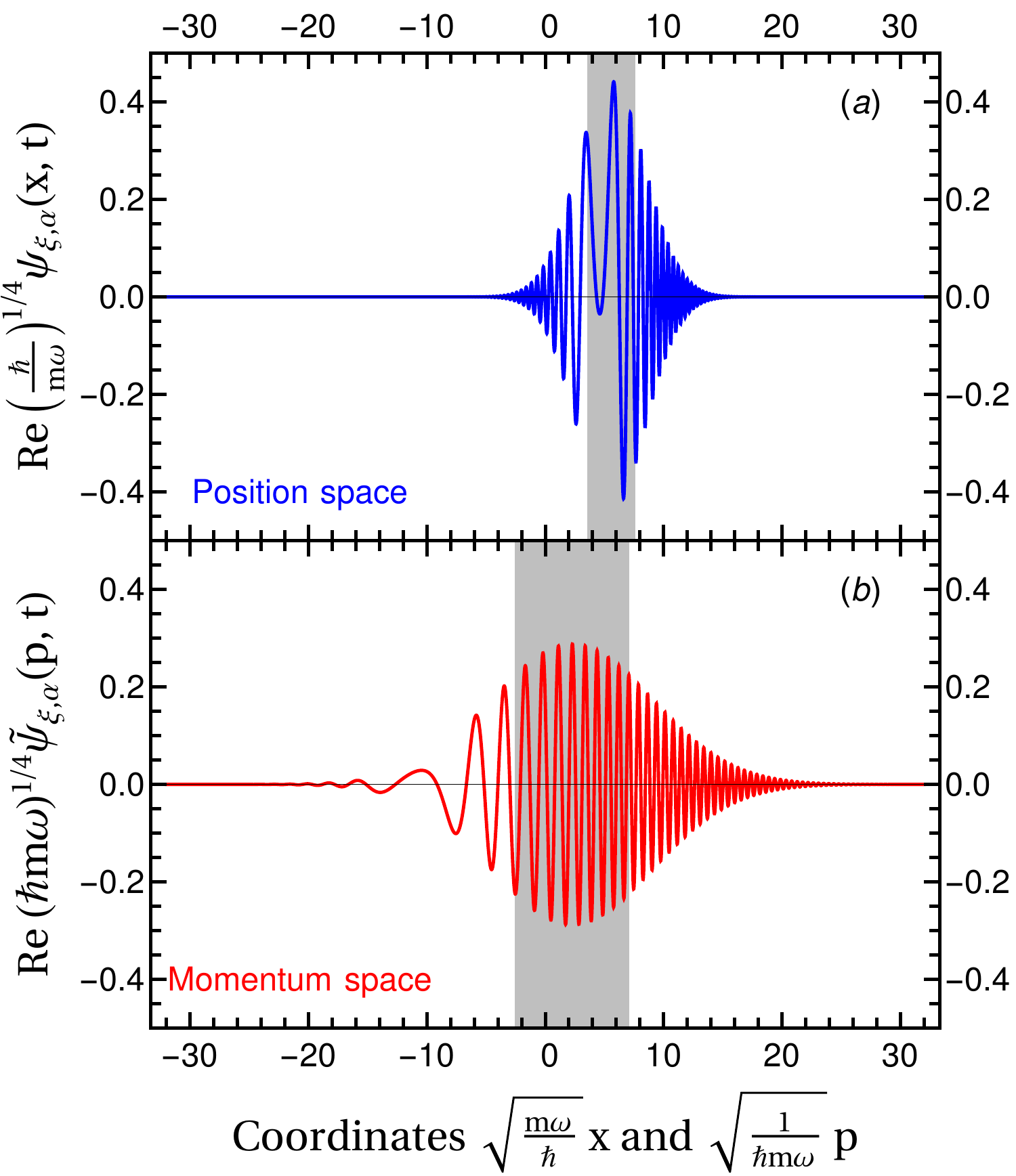}
\caption{%\href{https://pasteboard.co/JSeuCA5O.gif}{(\color{blue}Enhanced online)} 
(Enhanced online) [URL will be inserted by AIP] Snapshot of the wavefunctions of a squeezed coherent state with $r=2$, $\phi=0$ and $\alpha=3+3i$ ($x_0=3\sqrt{\frac{2\hbar}{m\omega}}$ and $p_0=3\sqrt{2m\omega\hbar}$) at the time instant $\omega t=0.4$; click enhanced online link for the full animation. Additional frames, which can be composed into a gif stack can be generated with the Mathematica notebook file available in the supplementary materials. The grey region indicates the fluctuations, and highlights the region extending from the average of the position or momentum by plus or minus the uncertainty in position or momentum. Both wavefunctions and the coordinates have their dimensions removed.\label{fig:wfs}}
\end{figure}

Even though the wavefunction is just a Gaussian, its time evolution and how it relates to uncertainty is interesting to visualize. We provide a Mathematica notebook in the supplementary materials\cite{supplemental} that allows one to compute and plot the position space and momentum space wavefunctions as a function of time along with a grey shaded region indicating the average and the fluctuations of position and momentum in the instantaneous wavefunction. In Fig.~\ref{fig:wfs}, we plot one typical panel that is output from the notebook (the animation can be seen by clicking the ``Enhanced online'' link). Note that the uncertainty product is minimal only when $\phi-2\omega t=0$ or $\pi$. At other times, the uncertainty product is not minimal and can become quite large. A measurement of the position or momentum coordinate has enhanced sensitivity (is squeezed) when the dimensionless variance is less than $\frac{1}{\sqrt{2}}$, which is what we have in a coherent state.

\section{Exercises for students}

In this section, we discuss some exercises that can be used with students working through this material. We suggest that one develops the wavefunctions either in the position-space basis or in the momentum-space basis, and then asks the students to work out the details and results in the other basis.

%Discuss different exercises---compute in position space, ask them to do momentum space

The overlap of two squeezed-coherent states appears to be extremely complicated when { examined} from the operator viewpoint. But, because one now has the wavefunctions in terms of Gaussians, the calculation is rather straightforward. In particular, labeling the squeezed coherent state as $|\xi,\alpha\rangle$, we have
\begin{equation}
    \langle \xi,\alpha|\xi^\prime,\alpha^\prime\rangle=\int_{-\infty}^\infty dx\,\langle \xi,\alpha|x\rangle\langle x|\xi^\prime,\alpha^\prime\rangle=\int_{-\infty}^\infty dx\, \psi_{\xi,\alpha}^*(x)\psi_{\xi^\prime,\alpha^\prime}(x).
\end{equation}
This is the integral of the product of two Gaussians; the product of two Gaussians is itself a Gaussian, so the integral can be immediately evaluated. Working out the coefficients is a bit tedious, so we do not work out the details here.

%Overlap of squeezed coherent states from Gaussian integrals

We employed two disentangling identities in this work (there is a third one discussed in the appendix needed for the momentum-space calculation with operators). These identities can be derived fairly quickly by simply factorizing two-by-two matrices, as we show in the appendix. Students can be asked to compute these identities. There is also a faithful three-dimensional representation of the symplectic group given by
\begin{equation}\label{eq:K_3d}
K_{+}=\sqrt{2}\begin{pmatrix}
0&1&0\\0&0&1\\0&0&0
\end{pmatrix}\quad
K_{-}=-\sqrt{2}\begin{pmatrix}
0&0&0\\1&0&0\\0&1&0
\end{pmatrix}\quad
K_{0}=\begin{pmatrix}
1&\phantom{-}0&\phantom{-}0\\0&\phantom{-}0&\phantom{-}0\\0&\phantom{-}0&-1
\end{pmatrix}.
\end{equation}
Students can verify the identity holds for the three-dimensional matrices after deriving it with the two-dimensional ones. This helps them learn that the identities really hold for any operators that satisfy the commutator relations.

%Verify disentangling identities

Finally, one can compute the time-dependent uncertainty relations, if desired, from the wavefunctions. In particular, one can show how the average position and momentum relate to the parameters $x_0$ and $p_0$, and one can compute the uncertainty product. Because time dependence is easy to incorporate afterwards, one can quickly go from the results at one time to results valid at all times.

\section{Relationship to LIGO}
LIGO employs a Michelson-Morley interferometer to measure small displacements corresponding to gravitational wave disturbances of the apparatus. In a Michelson-Morley interferometer, the beam splitter has two input ports and two output ports. Only one input port is used, however, in the standard design. Because light oscillates too rapidly for us to measure its electric field directly, the LIGO detectors use a balanced homodyne detection scheme, where a reference beam and the measured beam have the same frequency. {Homodyne detection is often thought to boost small signals by amplifying them due to the mixing with the high-powered reference beam, but the noise is also boosted, leading to no improvement of the signal to noise of the measured beam. But there often is an improvement of the signal relative to dark current noise in the photodetectors, which is not amplified by homodyning. Nevertheless,} the fluctuations in the quadrature variables (real and imaginary parts of the electric field) remain limited by the vacuum fluctuations on the unused input port. By instead inputting a squeezed-vacuum state on the second input port, one can reduce the fluctuations if the measurements are made precisely at the times where the particular quadrature variable has the smallest uncertainty (the uncertainty relations in Eqs.~(\ref{eq:x_unc}) and (\ref{eq:p_unc}) become time dependent with $\phi\to\phi-2\omega t$). It turns out that an enhancement of the sensitivity by a factor $\chi$ increases the size of the universe that can be observed by a factor of $\chi^3$. Hence, the current increase in sensitivity by about 30\% allows advanced LIGO to observe about twice as much of the universe than it can without employing the squeezed light. { One can see that the squeezed-coherent { wavefunctions} enter only tangentially to this analysis, as they can be employed to calculate the uncertainty relations of the measurement quadratures. But, they can still help the students visualize what squeezed states are and are thereby important in helping them better understand the details about how LIGO works.}

\section{Conclusions}

In this work, we showed three different ways that one can calculate the wavefunction of squeezed-coherent states of the simple harmonic oscillator. We believe that the success of LIGO should motivate instructors to introduce students to the concept of squeezed states to explain how using a squeezed vacuum on the unused input port of the LIGO interferometer can improve the measurements, which are otherwise limited by quantum uncertainty effects from the vacuum fluctuations on the input port. But, squeezed states can be intimidating. They involve complex operator expressions, and there is no simple way to convert those expressions into something that is easily visualizable. Because the squeezed states are also minimal uncertainty states at specific times during their time evolution, they must correspond to Gaussian wavefunctions in position space or momentum space. Since Gaussian wavefunctions should be more familiar to students, this provides a nice avenue to make connections with simpler concepts. We illustrate three different ways that one can derive these identities. The first is a differential equation based method. The most complicated aspect of that derivation is determining the overall complex phase of the wavefunction, which requires some nontrivial identities and operator algebra. The second way we did this involves using the wavefunctions of the simple harmonic oscillator to expand the squeezed-coherent state wavefunction. This is likely to be the most familiar, because students usually have seen the simple harmonic oscillator wavefunctions. But it requires a highly nontrivial identity relating summations over Laguerre polynomials to confluent hypergeometric functions to complete it. The final derivation is a complete operator-based approach. In classes that emphasize working with operators, students will already be exposed to the three required operator relations: Hadamard lemma, Weyl form of the Baker-Campbell-Hausdorff formula, and the exponential disentangling identity. In this case, nothing else is needed, but the derivation requires about ten steps to complete, and hence should be instructor-led.

We believe that this combination of creating sense of minimum uncertainty states through Gaussians and understanding how the squeezed-coherent wavefunctions relate to Gaussians is likely to aid students in learning these concepts. But we have not engaged in research to establish that this holds. We hope interested physics education researchers will explore this question (or related questions on this theme) in the future. In the meantime, we hope instructors will try using this in the classroom.

%provided three different strategies for including the squeezed state wavefunctions into a curriculum. This reinforces the concept of a minimum uncertainty state, that MUS are Gaussians and helps clarify just what a squeezed state really is.

\section{Supplementary materials}

Supplementary materials include a Mathematica notebook that allows one to compute gif stacks that visualize the time evolution of wavefunctions for different squeezing parameters, and the animation of the still image in the figure. [URL will be inserted by AIP] %\textit{For review purposes, the supplementary materials are available anonymously by clicking the following link:} {\color{blue}\href{https://gofile.io/d/924Imd/}{Supplementary materials.}}

\acknowledgments

We acknowledge funding from the National Science Foundation under grant number PHY-1915130. JKF also acknowledges support from the McDevitt bequest at Georgetown. We acknowledge useful discussions with K. Schonhammer, who helped us calculate the constant $c$ in the differential-equation-based approach. {We also acknowledge useful discussions with Carlton Caves, Michael Raymer and Val\'erian Thiel about experimental realizations of squeezed-coherent states in optical frequency combs, microscopy, measuring ion displacements in ion traps, and in microwave-cavity-based searches for axions (dark matter).}

\appendix

\section{Mathematical identities used in this work}

%Describe mathematical identities needed---Hadamard, BCH in Weyl form, exponential disentangling. Show how to derive exponential disentangling. Discuss Newton's binomial theorem and Erdelyi identity??

Most undergraduate quantum textbooks do not emphasize operator methods; they instead emphasize wavefunctions and differential equations. This is unfortunate, because it does not give students an opportunity to develop skill in working with operators. In this appendix, we present the different operator identities that we employ in this work, to ensure that it is clear how the calculations in the main text are completed.

The first operator identity we work with is the Hadamard lemma~\cite{hadamard}. As far as we can tell Hadamard had nothing to do { with} this, and it most likely originates with Campbell, but the name is often used, and it avoids conflation with the Baker-Campbell-Hausdorff formula, so we use it here.
The identity can be easily derived using a Taylor series expansion or via induction. It is
\begin{equation}
e^{\hat{A}}\hat{B}e^{-\hat{A}}=\hat{B}+[\hat{A},\hat{B}]+\frac{1}{2!}[\hat{A},[\hat{A},\hat{B}]]+\frac{1}{3!}[\hat{A},[\hat{A},[\hat{A},\hat{B}]]]+\cdots\label{eq:hadamard}
\end{equation}
where higher-order terms involve higher-order nested commutators (n-fold nesting for the nth term after the first). The formula is useful in two cases: (i) when the nested commutators truncate and (ii) when the nested commutators repeat in a pattern. Most usage in this work falls into category (i), but the computation of the time-evolution phase factors for the ladder operators follows from category (ii).

This results in the exponential re-ordering identity, { indeed} the equality
\begin{equation}
    e^{\hat{A}}(\hat{B})^n e^{-\hat{A}}=(e^{\hat{A}}\hat{B}e^{-\hat{A}})^n\label{eq:power_identity}
\end{equation}
allows us to move the Hadamard lemma similarity transformation \textit{inside} the arguments of functions. { If $f(\hat{B})$ has a Taylor-series expansion $f(\hat{b})=\sum_{n=0}^\infty f_n\hat{B}^n$, then we find that $e^{\hat{A}}f(\hat{B})=\sum_{n=0}^\infty f_n\left (e^{\hat{A}}\hat{B}^ne^{-\hat{A}}\right )e^{\hat{A}}$. Using Eq.~(\ref{eq:power_identity}) and resumming the series gives us}
\begin{equation}
    e^{\hat{A}}f(\hat{B})=f(e^{\hat{A}}\hat{B}e^{-\hat{A}})e^{\hat{A}}.\label{eq:braiding}
\end{equation}
The argument of the function can be evaluated by the Hadamard lemma. 

Next, we discuss the Baker-Campbell-Hausdorff formula\cite{baker,campbell,hausdorff,dynkin}, which we only need in the so-called Weyl form, which holds when both $[\hat{A},[\hat{A},\hat{B}]]=0$ and $[\hat{B},[\hat{A},\hat{B}]]=0$. { This can be derived by calculating the derivative of $e^{\lambda(\hat{A}+\hat{B})}e^{-\lambda\hat{B}}e^{-\lambda\hat{A}}$ with respect to $\lambda$, which gives us
\begin{eqnarray}
    \frac{d}{d\lambda}\left (e^{\lambda(\hat{A}+\hat{B})}e^{-\lambda\hat{B}}e^{-\lambda\hat{A}}\right )&=&e^{\lambda(\hat{A}+\hat{B})}(\hat{A}+\hat{B}-\hat{B})e^{-\lambda\hat{B}}e^{-\lambda\hat{A}}-e^{\lambda(\hat{A}+\hat{B})}e^{-\lambda\hat{B}}\hat{A}e^{-\lambda\hat{A}}\\
    &=&e^{\lambda(\hat{A}+\hat{B})}e^{-\lambda\hat{B}}\underbrace{e^{\lambda\hat{B}}\hat{A}e^{-\lambda\hat{B}}}_{\text{Hadamard lemma}}e^{-\lambda\hat{A}}-e^{\lambda(\hat{A}+\hat{B})}e^{-\lambda\hat{B}}\hat{A}e^{-\lambda\hat{A}}\\
    &=&e^{\lambda(\hat{A}+\hat{B})}e^{-\lambda\hat{B}}(\hat{A}+[\hat{B},\hat{A}]-\hat{A})e^{-\lambda\hat{A}}\\
    &=&[\hat{B},\hat{A}]\left (e^{\lambda(\hat{A}+\hat{B})}e^{-\lambda\hat{B}}e^{-\lambda\hat{A}}\right ),
\end{eqnarray}
after we use the fact that the derivative of the exponential of an operator can have the operator pulled down on either side of the exponential and the fact that the Hadamard lemma truncates after two terms in this case. The fact that the commutator $[\hat{B},\hat{A}]$ commutes with $\hat{A}$ and $\hat{B}$ means we can immediately integrate this expression to find that
\begin{equation}
    e^{\lambda(\hat{A}+\hat{B})}e^{-\lambda\hat{B}}e^{-\lambda\hat{A}}=e^{\frac{\lambda^2}{2}[\hat{B},\hat{A}]},
\end{equation}
because the operator on the left hand side is equal to the identity operator at $\lambda=0$. Finally, we set}
$\lambda=1$ {  and rearrange the identity} to yield
\begin{equation}
    e^{\hat{A}+\hat{B}}=e^{\hat{A}}e^{\hat{B}}e^{-\frac{1}{2}[\hat{A},\hat{B}]}\label{eq:BCH}
\end{equation}
and { $e^{\hat{A}+\hat{B}}e^{\frac{1}{2}[\hat{A},\hat{B}]}=e^{\hat{A}}e^{\hat{B}}$}, when the exponential of the commutator is moved to the left-hand-side.

Finally, we work with the exponential disentangling identity\cite{disentangling}; our approach uses the matrix factorization methodology.\cite{mufti} ~We discuss this in more detail than the other two, because it is less familiar and less used. There are many complex derivations of this identity (for example, an operator-based proof appears in advanced textbooks\cite{barnett_radmore} and in the literature\cite{dasgupta}), but if one can derive it in terms of the generators in a faithful representation, then it holds at the Lie-algebra operator level. So, all of these identities can simply be computed by factoring $2\times 2$ matrices. We explain in detail how this works for the three symplectic group identities needed in this work. The reason why the symplectic group is needed is that the three operators
\begin{equation}
    \hat{K}_0=\frac{1}{4}(\hat{a}^\dagger\hat{a}+\hat{a}\hat{a}^\dagger),~~\hat{K}_-=\frac{1}{2}\hat{a}^2,~~\text{and}~~\hat{K}_+=\frac{1}{2}(\hat{a}^\dagger)^2,\label{eq:koperators}
\end{equation}
satisfy the symplectic Lie algebra: $[\hat{K}_0,\hat{K}_\pm]=\pm\hat{K}_\pm$ and $[\hat{K}_+,\hat{K}_-]=-2\hat{K}_0$. A $2\times 2$ representation of this algebra is given by
\begin{equation}
    K_0=\frac{1}{2}\begin{pmatrix} -1&0\\\phantom{-}0&1\end{pmatrix},~~K_-=\begin{pmatrix}0&1\\0&0\end{pmatrix}\label{eq:kmatrices},~\text{and}~K_+=\begin{pmatrix}\phantom{-}0&0\\-1&0\end{pmatrix}.
\end{equation}
These matrices allow us to compute an exact result for their exponentials, because they square to a matrix proportional to the identity or to the zero matrix. We find that
\begin{equation}
    e^{cK_0}=\begin{pmatrix}\phantom{-}e^{-\frac{c}{2}}&0\\0&e^{\frac{c}{2}}\end{pmatrix},~~e^{bK_-}=\begin{pmatrix}1&b\\0&1\end{pmatrix},~~\text{and}~~e^{aK_+}=\begin{pmatrix}\phantom{-}1&0\\-a&1\end{pmatrix}.\label{eq:kexponentials}
\end{equation}
The derivation is possible because we can calculate the exponential of any linear combination of the $K$ matrices exactly. We then calculate ordered products of exponentials of the matrices in Eq.~(\ref{eq:kmatrices}). We illustrate this quickly for the three cases needed in this work.

The first example is $\exp(-\xi K_++\xi^* K_-)$. One can compute
\begin{eqnarray}
    \exp\left [ \begin{pmatrix}0&\xi^*\\\xi&0\end{pmatrix}\right ]&=&\begin{pmatrix}1&0\\0&1\end{pmatrix}+\begin{pmatrix}0&\xi^*\\\xi&0\end{pmatrix}+\frac{1}{2}\begin{pmatrix}0&\xi^*\\\xi&0\end{pmatrix}^2+\frac{1}{6}\begin{pmatrix}0&\xi^*\\\xi&0\end{pmatrix}^3+\cdots\nonumber\\
    &=&\begin{pmatrix}\cosh|\xi|&\frac{\xi^*}{|\xi|}\sinh|\xi|\\\frac{\xi}{|\xi|}\sinh|\xi|&\cosh|\xi|\end{pmatrix},\label{eq:matrix_exponential}
\end{eqnarray}
which is an exact result. We also can compute
\begin{equation}
    e^{aK_+}e^{cK_0}e^{bK_-}=\begin{pmatrix}e^{-\frac{c}{2}}&be^{-\frac{c}{2}}\\-ae^{-\frac{c}{2}}&e^{\frac{c}{2}}-abe^{-\frac{c}{2}}\end{pmatrix}.\label{eq:matrix_product}
\end{equation}
Now, we simply equate matrix elements to determine $a$, $b$, and $c$. We find that
\begin{equation}
    a=-\frac{\xi}{|\xi|}\tanh|\xi|,~~b=\frac{\xi^*}{|\xi|}\tanh|\xi|,~~\text{and}~~c=-2\ln\left (\cosh|\xi|\right ).\label{eq:disentangling1}
\end{equation}
One can check that with these choices we also have $e^{\frac{c}{2}}-abe^{-\frac{c}{2}}=\cosh|\xi|$, as is required for the factorization to be valid. Converting to operators (which is valid because of the Lie group theory discussion), we then have
\begin{equation}
    e^{-\frac{1}{2}\xi(\hat{a}^\dagger)^2+\frac{1}{2}\xi^*\hat{a}^2}=e^{-\frac{1}{2}\frac{\xi}{|\xi|}\tanh|\xi|(\hat{a}^\dagger)^2}e^{-\frac{1}{2}\ln\left (\cosh|\xi|\right )(\hat{a}^\dagger\hat{a}+\hat{a}\hat{a}^\dagger)}e^{\frac{1}{2}\frac{\xi^*}{|\xi|}\tanh|\xi|\hat{a}^2}.\label{eq:disentangling2}
\end{equation}

The next disentangling identity we work out is for 
\begin{equation}
\exp\left (-\frac{1}{2}e^{i\phi}\tanh r(\hat{a}^\dagger-\hat{a})^2\right )=\exp\left (-e^{i\phi}\tanh r \hat{K}_++2e^{i\phi}\tanh r\hat{K}_0-e^{i\phi}\tanh r\hat{K}_-\right ).\label{eq:disentangling3}
\end{equation}
We have
\begin{equation}
    e^{-e^{i\phi}\tanh r K_++2e^{i\phi}\tanh r K_0-e^{i\phi}\tanh r K_-}=\begin{pmatrix}1-e^{i\phi}\tanh r&-e^{i\phi}\tanh r\\e^{i\phi}\tanh r&1+e^{i\phi}\tanh r\end{pmatrix},\label{eq:disentangling4}
\end{equation}
because the matrix $-K_++2K_0-K_-$ squares to zero. Now, we use the result in Eq.~(\ref{eq:matrix_product}) to determine the constants $a$, $b$ and $c$ for this factorization, which are
\begin{equation}
    a=-\frac{e^{i\phi}\tanh r}{1-e^{i\phi}\tanh r},~~b=-\frac{e^{i\phi}\tanh r}{1-e^{i\phi}\tanh r}~~\text{and}~~c=-2\ln\left (1-e^{i\phi}\tanh r\right ).\label{eq:disentangling5}
\end{equation}
Then we find the operator factorization becomes
\begin{equation}
\exp\left (-\frac{1}{2}e^{i\phi}\tanh r(\hat{a}^\dagger-\hat{a})^2\right )=e^{-\frac{1}{2}\frac{e^{i\phi}\tanh r}{1-e^{i\phi}\tanh r}(\hat{a}^\dagger)^2}e^{-2\ln\left (1-e^{i\phi}\tanh r\right )}e^{\frac{1}{2}\frac{e^{i\phi}\tanh r}{1-e^{i\phi}\tanh r}\hat{a}^2}.\label{eq:disentangling6}
\end{equation}

The last operator identity we need is very similar to the one we just { found}. We simply change the sign in the quadratic. The $2\times 2$ matrix squares to zero again, so the result is also simple to compute. We find that
\begin{equation}
\exp\left (-\frac{1}{2}e^{i\phi}\tanh r(\hat{a}^\dagger+\hat{a})^2\right )=e^{-\frac{1}{2}\frac{e^{i\phi}\tanh r}{1+e^{i\phi}\tanh r}(\hat{a}^\dagger)^2}e^{-2\ln\left (1+e^{i\phi}\tanh r\right )}e^{\frac{1}{2}\frac{e^{i\phi}\tanh r}{1+e^{i\phi}\tanh r}\hat{a}^2}.\label{eq:disentangling7}
\end{equation}

\end{document}